\documentclass[final,1p,times]{elsarticle}
\usepackage[english]{babel}
\languageshorthands{english}
\usepackage{epsfig}
\usepackage{url}
\usepackage{multirow}
\usepackage{color}
\usepackage{mathtools}
\long\def\COMMENT#1\ENDCOMMENT{\message{(Commented text...)}\par}

\journal{Journal of Discrete Algorithms}
\begin{document}
\begin{frontmatter}

%%%%%%%%%%%%%%%%%%%%%%%%%%%%%%%%%%%%%%%%%%%%%%%%%%%%%%%%%%%%%%%%%%%%%%%%%
%\title{Flexible Interpolated-Binary Search over Sorted Sets}
\title{Adaptive Search over Sorted Sets}

\author[me]{Biagio Bonasera}

\author[uin]{Emilio Ferrara}
\ead{ferrarae@indiana.edu}

\author[me]{Giacomo Fiumara}
\ead{gfiumara@unime.it}

\author[mi]{Francesco Pagano}
\ead{francesco.pagano@unimi.it}

\author[me]{Alessandro Provetti\corref{ap}}
\ead{ale@unime.it}
\cortext[ap]{Corresponding author}

\address[me]{Dept. of Mathematics and Informatics, Univ. of Messina, V.le F. Stagno D'Alcontres 31, I-98166 Messina, Italy}

\address[uin]{Center for Complex Networks and Systems Research, School of Informatics and Computing. Indiana Univ., Bloomington, USA}

\address[mi]{Dept. of Informatics, Univ. of Milan, Via Comelico, 39. I-20135 Milan, Italy}

%%%%%%%%%%%%%%%%%%%%%%%%%%%%%%%%%%%%%%%%%%%%%%%%%%%%%%%%%%%%%%%%%%%%%%%%%%
\begin{abstract}
We revisit the classical algorithms for searching over sorted sets to introduce
an algorithm refinement, called Adaptive Search, that combines the good features
of Interpolation search and those of Binary search. 
W.r.t. Interpolation search, only a constant number of extra comparisons is introduced.
Yet, under diverse input data distributions our algorithm shows costs comparable to that of Interpolation search, i.e., $O(\log{\log{n}})$ while the worst-case cost is always in $O(\log{n})$, as
with Binary search. 
On benchmarks drawn from large datasets, both synthetic and real-life, Adaptive search scores better times and lesser memory accesses even than  Santoro and Sidney's Interpolation-Binary search.
\end{abstract}
%%%%%%%%%%%%%%%%%%%%%%%%%%%%%%%%%%%%%%%%%%%%%%%%%%%%%%%%%%%%%%%%%%%%%%%%%%

\end{frontmatter}
%%%%%%%%%%%%%%%%%%%%%%%%%%%%%%%%%%%%%%%%%%%%%%%%%%%%%%%%%%%%%%%%%%%%%%%%%%

%%%%%%%%%%%%%%%%%%%%%%%%%%%%%%%%%%%%%%%%%%%%%%%%%%%%%%%%%%%%%%%%%%%%%%%%%%%%%%%%
\section{Introduction}\label{sec:intro}
We revisit the classical algorithms for searching over sorted sets to introduce a new algorithm, called Adaptive search (AS), that combines the good features of Interpolation search and those of Binary search \cite{DBLP:book/algo/2009}. 

%%%%%%%%%%%%%%%%%%
%\begin{theorem}
\noindent
The membership problem can be formally defined as follows. 

\noindent 
\textbf{instance}: 

\begin{itemize}
	\item ${\cal S} = \left\{{a_1 ,a_2,...,a_n }\right\}$, a set of $n$ distinct, sorted elements,\\ with $a_i < a_{i+1}$ , $1\leq i \leq n-1$;  
	\item an element \textit{key}
\end{itemize}

\noindent 
\textbf{question}: Does \textit{key} belong to the set represented by ${\cal S}$ ($key\in{\cal S}$) ? 

\medskip
There exist two classical algorithms for searching over sorted sets: Binary search (BS) \cite{DBLP:book/algo/2009} and Interpolation search (IS) \cite{AndMat93}; both take advantage of the ordering of the instance to minimize the number of keys that must be accessed. 

In BS, the worst-case computational cost is $O(\log n);$ this result is independent of data distribution over the instance. 
Notice that in search the worst-case is rather important as it corresponds to an unsuccessful membership query.

Vice versa, the Interpolation Search algorithm is more efficient than BS when the elements of ${\cal S}$ are distributed uniformly or \textit{quasi-uniformly%
\footnote{By quasi-uniform data distribution we intended, informally, that the distance between two consecutive values of ${\cal S}$ does not vary much.
}} %endfootnote 
over the $[a_1, a_n]$ interval; the computational cost is in $O(\log \log n)$. 

Unfortunately, Interpolation search degrades to $O(n)$ when data is not uniformly distributed (in the sense above). 
This is particularly inconvenient when searching over indexes of large databases, where it is crucial to minimize the number of accesses%
\footnote{In this discussion we do not consider the advanced techniques, viz. the exploitation of locality, that underlie search over large database indexes.}. 

In this work we propose an algorithm, called Adaptive search (AS) that refines Interpolation search and minimizes the number of memory accesses needed to complete a search. 
AS is \textbf{adaptive} to the values by means of a {\em mixed} behavior: it combines the independence from the distribution of BS with the good average costs of IS. 
 
W.r.t. Interpolation search, AS requires only a constant number of extra comparisons. 
Yet, under several relevant input data distributions our algorithm shows average case costs comparable to those of interpolation, i.e., $O(\log{\log{n}}),$ while the worst-case cost remains in $O(\log{n})$, as
with Binary search. 

Comparison with a more recent literature is also encouraging: both on synthetic and real datasets AS has better times and lesser memory accesses than Santoro and Sidney's Interpolation--Binary search \cite{SanSid85}. 
Also, it is easier to implement and more broadly applicable that the approach of Demaine et al.\cite{DemJonPat04} to searching non-independent data.

%%%%%%%%%%%%%%%%%%%%%%%%%%%%%%%%%%%%%%%%%%%%%%%%%%%%%%%%%%%%%%%%%%%%%%%%%%%
\section{The Adaptive Search algorithm} 
Given an ordered set ${\cal S},$ allocated on an array \textit{A,} and an element \textit{key} that is searched, we define the following:

\begin{description}
	\item[{A[bot]:}] the minimum element of the subset (at the beginning, {\itshape bot} = 1);

	\item[{A[top]:}] the maximum element of the subset (at the beginning, {\itshape top} = $\left|{\cal S}\right|$);

	\item[{A[next]:}] interpolation element, i.e. what IS would choose, and

	\item[{A[med]:}] the el. halfway between {\it bot} and {\itshape top}, i.e., what BS would choose.
\end{description}

Our algorithm consists, essentially, of a while cycle. 
At each iteration, we consider ${\cal S} = \{A[bot],..,A[top]\}$ and we set:

\begin{eqnarray*}
next = bot + \left\lfloor \frac{key-A[bot]}{A[top]-A[bot]} *(top-bot)\right\rfloor
\end{eqnarray*}

\noindent 
Variable \textit{next} defined above contains the index value that bounds the array segment on which our AS algorithm will recur on. 
As with interpolation, the instance is now \textit{clipped:}

\begin{center}
\begin{math} 
{\cal S}^\prime = 
\left\{ 
\begin{array} {l}
\left\{A[bot],...,A[next]\right\} \  \mbox{if}\  A[bot]\leq key \leq A[next]\\
\\
\left\{A[next],...,A[top]\right\} \  \mbox{otherwise}%se\  A[pos]< key \leq A[sup]
\end{array}
\right.
\end{math}
\end{center}

\noindent 
To do so, we set the new boundaries of the segment containing $\mathcal{S}^\prime$:

\begin{eqnarray*}
	\left\{ 
	\begin{array} {l}
		top = next\   \mbox{if}\  A[bot]\leq key \leq A[next]\\
		\\
		bot = next\  \mbox{otherwise} %se\  %A[pos]< key \leq A[sup]
	\end{array}
	\right.
\end{eqnarray*}

\noindent 
The computation is now restricted to the segment that would have been considered by IS.
Next, the median point is computed over such restricted segment, \textsl{rather than on the whole input. }
Vice versa, if interpolation returns a shorter interval than BS would have, we keep the result of the interpolation step:

\begin {tabbing}
if $\left|{\cal S}^\prime \right| > \frac{\left|{\cal S}\right|}{2}$ \= then \= $next = med = bot+ \frac{top-bot}{2}$;\\
 \> elseif $key = A[next]$ then \textit{key} is found and we terminate;\\
 \> \>                     elseif \= $key > A[next]$ then $bot=next+1$;\\
 \> \> \>                          else $top=next-1$ (must be $key < A[next]$).
\end {tabbing} 

\noindent
At the end of the iteration, ${\cal S}^\prime  = \left\{A[bot],...,A[top]\right\}$, and, clearly, $\left|{\cal S}^\prime \right|< \frac{\left|{\cal S}\right|}{2}$. 
Finally: 

\begin{tabbing}
if $A[bot]<key<A[top]$ \= then iterate search on ${\cal S}^\prime;$\\ 
                       \> else $key \notin {\cal S}$ and we terminate with \textit{no.}
\end{tabbing}

\noindent
From the point of view of computational costs, we could summarize the following: our algorithm may spend up to double number of operations than IS in carefully finding out the best halving of the search segment, which in turn will mean that less iterations shall be needed to complete. 
%%%%%%%%%%%%%%%%%%%%%%%%%%%%%%%%%%%%%%%%%%%%%%%%%%%%%%%%%%%%%%%%%%%%%%
%\subsection {Computational cost of Adaptive Search}
By means of standard cost analysis techniques, we have the following results:

\begin{itemize}
\item Best case: \textit{key} is found, with a constant number of comparisons: $\Theta(1)$;

\item Worst case: the intervals between values are unevenly distributed; hence, the interval found by the BS technique is always the shortest. 
As a result, AS will execute essentially the same search as BS, with equal $O(\log n)$ time complexity (but more operations at each level), and

\item Average case: we consider the average case to be when the distance between two consecutive values of $\mathcal{S}$ varies according to the normal (Gaussian) probability distribution.  
In such cases, AS executes exactly as IS so its cost is in $O(\log \log n).$ 

To see why this is the case, consider the probability that a given input (an ordered set in array A, and a value key to be search) elicit the IS case under very mild assumptions about the distribution of the ordered values in A. 
For any given instance, we fix $N=bot-top$ as the number of \textit{gaps} between two consecutive values of the ordered set ${\cal S}$. 
Also, we fix $R=A[top]-A[bot]+1$ as the numeric \textit{range} over which the input values appear% 
\footnote{Whenever key is out of such range, i.e., $key <A[bot]$ or $key > A[top]$ we exit with 'no' in constant time, and so do other algorithms.}.  
It is reasonable to associate the input value \textit{key} to a random numerical variable $X_{key}$ distributed uniformly over the values interval $A[bot]\dots A[top].$

Now, consider N distinct random variables $X_1,\dots X_N$ representing the width of each gap between two consecutive values of ${\cal S}$: $X_i= A[i+1]-A[i]$.  
We associate each random variable $X_i$ with a normal probability distribution centered over $\frac{R}{N},$ given that $\sum_{i=1}^N X_i=R.$
Hence, the expected value of each $X_i$ is $\frac{R}{N}.$ 
Moreover, thanks to the property of the normal distribution by which the sum of normal distributions is a normal distribution itself, we can obtain that $Pr[\sum_{i=1}^{\frac{N}{2}} X_i\leq \frac{R}{2}]\leq \frac{1}{2}.$ 
We can then conclude that the average case, captured by the normal distribution of the gaps, safely fails in the IS case.
\end{itemize}

%%%%%%%%%%%%%%%%%%%%%%%%%%%%%%%%%%%%%%%%%%%%%%%%%%%%%%%%%%%%%%%%%%%%%
\section{Relation with literature}
Only after our solution was conceived and implemented, have we become aware of an earlier work by Santoro and Sidney \cite{SanSid85} who devised a similar solution that combines (but does not \textsl{blend)} together interpolation and binary search. 
Although the asymptotic complexity is the same, there are some marked differences between their solution and ours, let's discuss them now.

Santoro--Sidney's algorithm, called Interpolation-Binary search, is based on the idea that interpolation search is useful, from the point of view of costs, only when the array searched is larger than a given threshold. 
When the considered array segment is smaller than a user-defined threshold, binary search is applied unconditionally.
Vice-versa, above the threshold an interpolation search step is applied, followed eventually by a binary search step. 

Unlike IBS, our algorithm makes, at each level of its iteration, a choice about which \textit{clipping} of ${\cal S}$ to apply. 
Hence, it is possible to show that for any input AS will not take more elementary operations than IBS. 

We have sought a statistical confirmation of this fact by running a set of experiment over random-generated ordered sets; the results are presented in detail in the next section.%
\footnote{The instances and the test times are available from the companion Web site.} 
We limited the testing of IBS to queries with parameter $\theta=2,$ which the authors suggested would work best.
For all parameter settings and for all data distributions considered AS outperformed IBS albeit the difference could sometimes be statistically insignificant. 

Two other works that address search over sorted sets have considered slight variations of the specification, that of Melhlhorn and Tsakalidis \cite{MehTsa93} and that of Demaine et al. \cite{DemJonPat04}. 
The former considered an extended data structure, the Interpolation Search Tree (IST) to optimize the dictionary operations, not just search, over the sorted set. 
As such, their solution is not comparable to ours as it seeks to optimize insertion and deletion times rather than speed up search. 

%%%%%%%%%%%%%%%%%%%%%%%%%%%%%%%%%%%%%%%%%%%%
%\subsection{The analysis of Demaine et al.}
%In this Section we summarize the results of 

The latter, i.e., Demaine's interpolation search for \textit{non independent} data is also not directly comparable to our work, but deserves a careful analysis.  
They define a deterministic metric of ``well-behaved'' or \textit{smooth} data that enables searching along the lines of interpolation search. 
Specifically, they define 

\begin{eqnarray*}
\Delta = \frac{max(x_i-x_{i-1})}{min(x_i-x_{i-1})}
\end{eqnarray*}

\noindent
i.e., the ratio between the largest and smallest \emph{gap} between two adjacent elements of ${\cal S},$ as the key parameter in measuring the well-behavedness of the input. 
A data structure is needed that maintains a dynamic dataset, that evenly divide the interval $(x_1,...,x_n)$ into $n$ \textit{bins,} named $B_1,\dots B_n$; each of them represents a range of size $\frac{x_n-x_1}{n}$.

Each bin $B_i$ stores in a balanced Binary Search Tree (BST) its elements, \textit{plus} the nearest neighbors above and below that set. 
Hence, searching for an element \textit{key} proceeds by interpolating on \textit{key} to find which $B_i$ it may lay in, i.e., 

\begin{eqnarray*}
i = \frac{(\mathit{key}-x_1)}{(x_n-x_1)}
\end{eqnarray*}

\noindent
then performing a search in the BST associated to $B_i.$ 
For their solution, Demaine et al. prove the following results:

\begin{itemize}
\item the worst-case search time is $O(\log n)$ and thus $O(\log \min\{\Delta,n\}),$ and 

\item the algorithm reproduces the $O(\log\log n)$ performance of interpolation search on data drawn independently from the uniform distribution.
\end{itemize}

%%%%%%%%%%%%%%%%%%%%%%%%%%%%%%%%%%%%%%%%%%%%%%%%%%%%%%%%%%%%%%%%%%%%%%%%%%%%%%%%%%%
\section{Experimental validation}
We have implemented AS, along with the other algorithms mentioned so far, in order to test its efficiency, on real data, vis-\`a-vis those in the literature. 
The testing platform consists of a Java implementation running on a PC with JRE 1.7, Windows 2003 server R2, dual Opteron CPU with 4GBs of RAM. 
The tests consisted of running a number of searches corresponding to 1/1000 of the size of the dataset; keys where randomly chosen, with at least 80\% of them successful. 
The results were normalized w.r.t. the number of queries.

%%%%%%%%%%%%%%%%%%%%%%%%%%%%%%%%%%%%%%%%%%%%%%%%%%%%%%%%%%%%%%%%%%%%%%%%%%%%%%%%%%%
\subsection{Validation across distributions}
As a first step, we considered random-generated benchmark instances (ordered arrays) of Java double data type, double-precision 64-bit IEEE 754 floating point values. 
Instances were randomly generated, with the following distribution types%

\begin{enumerate}
	\item \emph{uniform sparsity:} the gap between two consecutive values is fixed across the instance. 
	As unrealistic as it is, this case is useful in assessing whether AS introduces overheads.
	
	\item \emph{increasing sparsity:} the gap is actually growing, so the elements towards the end (i.e., the highest integer values) are more distant from each other than those at the beginning.
	
	\item \emph{stepwise sparsity:} the instance has zones with distinct, but fixed, gap sizes; the gap size grows towards the end of the array.
	
	\item \emph{Paretian:} the ``80--20'' rule applied to the summation of the values inside the instance, i.e., the summation of the first 80\% of elements is equal to the summation of the last 20\%.
	
\end{enumerate}

For each parameter setting we generated and tested 10 random instances, then computed the average. 
Also, values are normalized w.r.t. the number of queries, so as to make them comparable across instance sizes.  
The results, presented in Table \ref{tab:synthetic} compare the number of accesses, iterations and times of the four algorithms we considered.

\begin{table}[htb]
\begin{center}
$
\begin{array}[htb]{||c|c||r|r||r|r||r|r||}
\hline \hline
             &       & \multicolumn{2}{|c|}{\rm{no. of Accesses}}  & \multicolumn{2}{|c|}{\rm{no. of Iterations}} & \multicolumn{2}{|c|}{\rm{Times}} \\
				 &  \rm{Sizes:}     & 10^5 & 10^6 & 10^5 & 10^6 & 10^5 & 10^6 \\ 
\hline
\rm{Sparsity} & \rm{Algo.}     &      &      &      &      &      & \\
\hline
\multirow{4}{*}{uniform}   & BS  & 14,728 &  18,467 &  15,790 &  19,155 &  4.074,941 &    756,956 \\
                           & IS  &  4,743 &   4,919 &   2,888 &   3,068 &  2,553,635 &    700,328 \\
                           & IBS2& 19,644 &  23,397 &  26,922 &  33,613 & 25,273,348 &  3,639,200 \\
                           & AS  &  6,054 &   6,290 &   2,887 &   3,065 &  3,150,028 &  1,004,239 \\
\hline
\multirow{4}{*}{increasing}& BS  & 14,741 &  18,479 &  15,828 &  19,156 &    462,659 &    831,015 \\
                           & IS  & 19,613 &  26,619 &  18,906 &  25,978 &    948,907 &  2,994,142 \\
                           & IBS2& 19,338 &  23,502 &  22,581 &  29,177 &  1,345,126 &  3,499,578 \\
                           & AS  & 11,198 &  12,160 &   5,460 &   6,016 &    596,080 &  1,744,790 \\
\hline
\multirow{4}{*}{stepwise}  & BS  & 14,795 &  18,505 &  15,957 &  19,171 &    445,794 &    753,501 \\
                           & IS  &232,945 & 329,222 & 256,465 & 351,515 & 10,386,056 & 35,041,154 \\
                           & IBS2& 20,304 &  24,202 &  24,777 &  31,096 &  1,485,665 &  3,604,115 \\
                           & AS  & 12,055 &  12,968 &   6,129 &   6,708 &    652,009 &  1,505,453 \\
\hline
\multirow{4}{*}{Paretian}  & BS  & 14,793 &  18,476 &  15,917 &  19,157 &    457,496 &    916,074 \\
                           & IS  & 17,536 &  21,702 &  16,028 &  20,209 &    839,989 &  2,768,519 \\
                           & IBS2& 20,252 &  24,180 &  25,339 &  31,904 &  1,509,791 &  3,900,253 \\
                           & AS  & 10,338 &  11,003 &   5,097 &   5,536 &    564,157 &  1,516,632 \\
\hline \hline
\end{array} $	
\end{center}
\caption{Averaged and normalized benchmark values over random instances with distinct data distributions. Times are in milliseconds.}
\label{tab:synthetic}
\end{table}
%%%%%%%%%%%%%%%%%%%%%%%%%%%%%%%%%%%%%%%%%%%%%%%%%%%%%%%%%%%%%%%%%%%%%%
%\input{../ICALP06/benchmarks/as-vs-ibs-05} % tables comparing 
%%%%%%%%%%%%%%%%%%%%%%%%%%%%%%%%%%%%%%%%%%%%%%%%%%%%%%%%%%%%%%%%%%%%%%

On aggregate, AS outperformed IBS2 as well, albeit the difference could sometimes be statistically insignificant. 
The distributions were designed to stress-test AS in an unfavorable setting, where quicker implementations of BS could easily make up for the extra number of iterations. 
Even though on uniform- and increasing-sparsity instances Binary search can still run slightly faster than AS, on aggregation AS yields a huge advantage over all other algorithms, especially in terms of number of accesses and iterations.

\COMMENT
Surprisingly, the case for $\theta=2$ which the authors suggest as the best choice of the parameter actually yields the worst efficiency vis-\`a-vis AS. 
Table \ref{tab:as-vs-ibs2-increasing} compares IBS with $\theta=2$ with AS over instances with 
increasing sparsity, benchmark where IBS performed best. 
Even on the increasing-sparsity benchmark AS outperformed IBS at least 1 to 2 (accesses to the dataset). 

%%%%%%%%%%%%%%%%%%%%%%%%%%%%%%%%%%%%%%%%%%%%%%%%%%%%%
\input{../ICALP06/benchmarks/as-vs-ibs-2} % tables comparing 
%%%%%%%%%%%%%%%%%%%%%%%%%%%%%%%%%%%%%%%%%%%%%%%%%%%%%
\ENDCOMMENT

The full benchmark results and the source codes (in Java) will be made available on a dedicated Web site%
\footnote{\url{http://informatica.unime.it/adaptive-search/}}.

\COMMENT
%%%%%%%%%%%%%%%%%%%%%%%%%%%%%%%%%%%%%%%%%%%%%%%%%%%%%%%%%%%%%%%%%%%%%%%%%%%%%%%%%%%
\subsection {Cache awareness}
In our experiments we have focused on counting the number of accesses needed to solve the search problem rather than on the raw computation times. 
In particular, we assumed a context where the time to access the memory is many times higher than that needed for simple additions and comparisons. 
This setting is of fundamental importance to study the performance of this class of algorithms for applications to arrays distributed over a network, or indexes of large database).  
Therefore, we compared the three algorithms at hand by counting  the number of accesses to memory each one needed.

In the second experimental phase, we considered a \emph{caching mechanism.} 
In such a case, only the first access to each element is relevant, since subsequent accesses will be resolved at the cache level.
On the other hand, every search implies moving several contiguous elements to the cache at once (searching for the i-th element, involves bringing to cache $A[i-1]$, $A[i]$ and $A[i+1]$). 
In both IS and AS we do not count the access to the first and the last element of the array as we assume that boundaries change seldom.

To carry out a comparative test for the caching environment, we have resorted to a simulation. 
That is, we have simulated the workings of the cache by a function that works like this:

-it returns, in one step, the value the element of the array;

-it checks a flag to see whether the element -or one of its neighbors- has
been accessed earlier

-if not, then the function flags the element, its adjacent elements and finally 
 increases the access counter.
 
Therefore, at the end of the execution the access counter reflects the number of times we would have to access the external (slower) memory. 
Notice that the first and last element are always considered to be in the cache when computation begins.

Having to flush the cache at the start of every search, the computing time of our simulation would be significantly long. 
Therefore, to reduce testing times we choose to run only 100 searches for each instance.
\ENDCOMMENT

%%%%%%%%%%%%%%%%%%%%%%%%%%%%%%%%%%%%%%%%%%%%%%%%%%%%%%%%%%%%%%%%%%%%%%%%%%%%%%%%%%%%%%
\subsection{A real-life benchmark dataset}
To perform our analysis on real-life data with mixed or alternating distributions we used a public dataset on Facebook friendship released by Gjoka et al. \cite{gjoka2011practical}; it contains a graph of about 957 thousands vertices (each representing a user) and 58.4 millions edges (each representing a friendship relation). 
Since each user is identified by a unique integer, and the dataset is ordered by user-id, it represents an ideal benchmark for testing our Adaptive search algorithm as it gives to us one instance of about 1 million ordered integers. 
Also, the dataset can be split up in 9 distinct subinstances of 100k elements each. 
The collected user-ids depend on several factors and human intervention, e.g., users \textit{leaving} Facebook and thus having their ids removed, so subinstances turn out to have distinct data distributions. 
On top of that, the gaps between two consecutive user-ids depend also on how the sample was collected, as discussed, e.g., in \cite{DFFP11improving}. 
In other words, this real-world dataset arguably summarizes the bias introduced by user activity and web data collection.

To confirm these intuitions, we performed a statistical analysis of the distribution of the gaps w.r.t. a \textit{null model,} i.e., a random instance that we generated with the same size, the same range of the values and with gaps having the same average and standard deviation as the real dataset.
How would the gaps (and therefore the user-ids) in the Facebook dataset relate to their randomized version? 
The resulting null model instance turned out to have a very different distribution of the gaps.
For the whole dataset we found a Spearman's rank correlation coefficient equal to $4.95 \cdot 10^{-5}$; also Pearson's correlation coefficent was very low, at $1.66 \cdot 10^{-4}$; this indicates that the FB dataset is \textit{nonuniform.} 

We used the same platform and the same set-up as before for the testing; the first test considered the whole Gjoka's dataset and the aggregated results (averaged over 10 runs) are in Table \ref{tab:whole}. 
As per the synthetic benchmarks, we ran a number of searches corresponding to 1/1000 of the size of the dataset; keys where randomly chosen, with at least 80\% of them successful. 

\begin{table}[htb]
\begin{center}
$
\begin{array}[htb]{||c|r|r|r||}
\hline \hline
Algorithm & Accesses & Iterations & Time (ms)\\
\hline
BS & 18,439 & 19,136 & 6,236,787 \\

IS & 501,346 & 499,730 & 74,035,808 \\

IBS 2 & 24,097 & 31,474 & 28,205,959 \\

AS & 8,349 & 4,044 & 4,791,845\\
%ANS & 8,333 & 3,996 & 5.145,173\\
%AMS & 8,349 & 4,042 & 5.087,715 \\
%TS & 22,841 & 12,514 & 8.989,054 \\
\hline \hline
\end{array} $	
\end{center}
\caption{Benchmarks values over Gjoka's dataset}
\label{tab:whole}
\end{table}

Subsequently, we have sought to confirm these results over similar datasets having diverse value distributions. 
To do so, we repeated the test on 9 sub-instances of Gjoka's, each corresponding to 100k consecutive keys, i.e., positions (not values) 0--99.999, 100.000--199.999 and so on.  
In fact, the $L_2$ (Euclidean) distance from a uniform distribution of \textit{gaps} between two consecutive values, varies widely. 
Nevertheless, our AS algorithm performed well on each subset, as it is reported in Table \ref{tab:accesses-piecemeal}. 

\begin{table}[htb]
\begin{center}
$
\begin{array}[htb]{||c|r|r|r||r|r|r||r|r|r||}
\hline \hline
Instance & 1      & 2      & 3      & 4      & 5      & 6     & 7      & 8       & 9\\
\hline
IS    & 1,662  &  1,473 &  1,494  &  1,463  &  1,488  &  1,470  &  1,483  &  1,487  &  1,489 \\
BS    &   621  &   522  &  3,623  &    300  &    300  &    300  &    300  &    300  &    300 \\
IBS   & 2,177  &  2,028 &  1,951  &  2,044  &  2,043  &  2,051  &  2,057  &  2,047  &  2,068 \\
AS    &   889  &   711  &    860  &    307  &    310  &    311  &    309  &    309  &    309 \\
%1     &   893  &   713  &    862  &    307  &    310  &    311  &    309  &    309  &    309 \\
%2     &   889  &   711  &    860  &    307  &    310  &    311  &    309  &    309  &    309 \\
%3     & 2,118  &  1,861 &  1,870  &  1,854  &  1,857  &  1,894  &  1,873  &  1,871  &  1,842 \\
\hline \hline
\end{array} $	
\end{center}
\caption{Memory accesses over 9 subinstances of Gjoka's dataset}
\label{tab:accesses-piecemeal}
\end{table}

\begin{table}[htb]
\begin{center}
$
\begin{array}[htb]{||c|r|r|r|r|r|r|r|r|r||}
\hline \hline
Instance no. & 1      & 2       & 3       & 4       & 5       & 6       & 7       & 8       & 9 \\
\hline
IS    & 1,783  &  1,567  &  1,613  &  1,570  &  1,602  &  1,578  &  1,589  &  1,593  &  1,605 \\
BS    &   422  &    351  &  3,454  &    100  &    100  &    100  &    100  &    100  &    100 \\
IBS   & 2,834  &  2,633  &  2,466  &  2,630  &  2,618  &  2,636  &  2,644  &  2,643  &  2,661 \\
AS    &   417  &    355  &    453  &    100  &    100  &    100  &    100  &    100  &    100 \\
%         421  &   351  &   454  &   100  &   100  &   100  &   100  &   100  &   100
%         417  &   355  &   453  &   100  &   100  &   100  &   100  &   100  &   100
%       1,202  &  1041  &  1048  &  1043  &  1019  &  1066  &  1042  &  1044  &  1025
\hline \hline
\end{array} $	
\end{center}
\caption{Iterations over 9 subinstances of Gjoka's dataset}
\label{tab:iter-piecemeal}
\end{table}

\begin{small}
\begin{table}[htb]
\begin{center}
$
\begin{array}[htb]{||c|r|r|r|r|r||}%|r|r|r|r
\hline \hline
Instance\ no. & 1       & 2           & 3           & 4              & 5       \\%& 6          & 7          & 8       & 9 \\
\hline
IS    & 1,180,339  &    433,851  &    474,331  &    447,672  &   453,487  \\%&   453,342  &   481,395  &   71,925  &   70,068 \\
BS    &   682,630  &    252,233  &  1,862,590  &    126,926  &   127,868  \\%&   127,822  &   131,044  &  136,434  &  130,819 \\
IBS   & 5,104,407  &  2,600,980  &  2,533,918  &  2,723,071  &  2,523,096  \\%&  2,541,275  &  2,588,292  &  242,335  &  254,291 \\
AS    &   427,276  &    326,454  &    393,219  &    135,930  &   135,005  \\%&   134,669  &   136,642  &  141,063  &  140,683 \\
%437  &  353967  &  432723  &  164089  &  164637  &  166544  &  167331  &  170971  &  174948
%416,049  &  355002  &  424898  &  168330  &  169098  &  168434  &  188456  &  178116  &  169560
%735,502  &  624733  &  630146  &  659559  &  636049  &  654601  &  674824  &  656617  &  661510
\hline \hline
Instance\ no. & 6           & 7           & 8           & 9 &  \hbox{Sum} \\
\hline
IS       &    453,342  &    481,395  &   71,925  &   70,068 &  4,066,410 \\
BS       &    127,822  &    131,044  &  136,434  &  130,819 &  3,578,366 \\
IBS      &  2,541,275  &  2,588,292  &  242,335  &  254,291 & 21,111,665 \\
AS       &    134,669  &    136,642  &  141,063  &  140,683 &  1,970,941 \\
%437  &  353967  &  432723  &  164089  &  164637  &  166544  &  167331  &  170971  &  174948
%416,049  &  355002  &  424898  &  168330  &  169098  &  168434  &  188456  &  178116  &  169560
%735,502  &  624733  &  630146  &  659559  &  636049  &  654601  &  674824  &  656617  &  661510
% sums for the remaining 3 versions are: 2232210, 2237943, 5933541
\hline \hline
\end{array} $	
\end{center}
\caption{Times (in milliseconds) for search over 9 subinstances of Gjoka's dataset}
\label{tab:times-piecemeal}
\end{table}
\end{small}

%\noindent
%At this point of our validation, we can rule out any specific \emph{positive bias} of the benchmark. 

%%%%%%%%%%%%%%%%%%%%%%%%%%%%%%%%%%%%%%%%%%%%%%%%%%%%%%%%%%%%%%%%%%%%%%%%%%%%%%%%%%
\section{Conclusions} 
Even though we have considered only the simplest instance of search, i.e., ordered sets of integers, it turns out that this case is of great practical interest when we consider large dataset extracted from, e.g., crawling Web pages or Online Social Networks, where users/resources are identified by simple integer keys. 
This is notably the case with Facebook, which assign to each subscriber a user-id consisting of a \textit{progressive} integer. 
On such type of data, our solution shows a marked improvement over the literature. 
The results of experiments described in the previous section lead us to draw the following conclusions:

\begin{enumerate}
	\item The performances of our AS algorithm vis-\`a-vis those IS and BS are very good and improve as \textit{n} grows;

	\item The number of accesses needed by AS is less than those of BS. 
	The cost analysis of IS suggests that on certain instances, i.e., when sparsity grows, our algorithm needs between $\log{n}$ and $2\log{n}$ accesses. 

	\item Our method for selecting the search interval succeeds in preventing the irregularities of data distribution from affecting performances; indeed, the number of accesses required remains $\cong$ $\log{\log{n}}$.

	\item While the asymptotic complexity of our AS algorithm is the same as Santoro's IBS, we have found that -on relatively diverse benchmarks- AS often needs half or less of the memory accesses than IBS. 
	
	\item Even though we could not yet run a complete study on large datasets, we have indication that the results presented here are likely to be confirmed for search dictionaries (considered by \cite{MehTsa93,AndHagHasPet01}).
\end{enumerate}

\noindent
An interesting open question is whether instances that elicit the worst case ($2\log n$ comparisons) for AS can actually be found, and how likely they are to appear within real datasets. 

%%%%%%%%%%%%%%%%%%%%%%%%%%%%%%%%%%%%%%%%%%%%%%%%%%%%%%%%%%%%%%%%%%%%%%%%%%%%%%%%%%
\section*{Acknowledgments}

\noindent
Thanks to Minas Gjoka for making the dataset studied in this work available. 
%Our work benefited from useful discussions and suggestions from our colleagues P. Boldi, F. Bassetti, C. Mereghetti and C. Rudipelli.

%%%%%%%%%%%%%%%%%%%%%%%%%%%%%%%%%%%%%%%%%%%%%%%%%%%%%%%%%%%%%%%%%%%%%%%%%%%%%%%%%%%%%%%%%
\bibliographystyle{elsarticle-num}
\bibliography{adaptive-search}
%%%%%%%%%%%%%%%%%%%%%%%%%%%%%%%%%%%%%%%%%%%%%%%%%%%%%%%%%%%%%%%%%%%%%%%%%%%%%%%%%%%%%%%%%
%%%%%%%%%%%%%%%%%%%%%%%%%%%%%%%%%%%%%%%%%%%%%%%%%%%%%%%%%%%%%%%%%%%%%%%%%%%%%%%%%%%%%%%%%
\end{document}